# A globally-applicable disease ontology for biosurveillance; Anthology of Biosurveillance Diseases (ABD)

A.R. Daughton*, R. Priedhorsky, G. Fairchild, N. Generous, A. Hengartner, E. Abeyta, N. Velappan, A. Lillo, A. Deshpande
Los Alamos National Laboratory
Los Alamos, NM, USA
*adaughton@lanl.gov

K. Stark
Digital Infuzion, Inc.
Gaithersburg, MD, USA

*Abstract*—Biosurveillance, a relatively young field, has recently increased in importance because of its relevance to national security and global health. Databases and tools describing particular subsets of disease are becoming increasingly common in the field. However, a common method to describe those diseases is lacking. Here, we present the Anthology of Biosurveillance Diseases (ABD), an ontology of infectious diseases of biosurveillance relevance.

## I. Introduction

Biosurveillance is a relatively young field. While the first health surveillance systems are from the fourteenth and fifteenth centuries during the Black Death (a large outbreak of pneumonic plague) [1], health surveillance was only recognized as its own field in the 1960s [1], and the U.S' first national strategy for biosurveillance was released only in 2012 [2]. Further, this discipline is broad in nature. The national strategy for biosurveillance calls for systems to "detect, track, investigate, and navigate incidents affecting human, animal, and plant health, thereby better protecting the safety, well-being, and security of the American people" [2], but biosurveillance often falls under global health security [3]. Because of the breadth that human, plant and animal health encompasses, only recently has there begun to be consensus in the field about what the full "biosurveillance" spectrum is, what data streams are included in such surveillance [4], and further, what diseases are relevant.

Los Alamos National Laboratory provided one of the first comprehensive analyses of biosurveillance goals and data streams and used that analysis to implement the Biosurveillance Resource Directory (BRD) [4]. During development of this database, it became clear that appropriate description of biosurveillance resources required an unambiguous description of relevant diseases. However, existing diesease ontologies describe particular populations (e.g. animals, but not humans, or visa versa), and tend to rely on clinical characteristics that may or may not be applicable to systems sureying various global domains (e.g. plants) or using methods that avoid clinical diagnosis of disease. To that end, we developed a globally applicable ontology for biosurveillance application named the "Anthology of Biosurveillance Diseases" (ABD).

## II. Ontology Development

### A. Logistic Ontology Requirements

It is important that the developed ontology meet the following requirements:

- *Correctly identify diseases from synonyms:* "German measles", for example, is not a term for measles, but rather for the disease rubella. Similarly, "rubeola" refers not to rubella, but to measles [5]. It was vital to ensure that our ontology capture these synonyms, and others like them, without confusion.

- *Describe organisms associated with diseases, either by causing a disease, spreading a disease, or being infected with a disease at varying levels of resolution:* Some diseases, such as dengue and chikungunya, are spread by specific vectors, in this case *Aedes aegypti* and *Aedes albopictus* [6]. Other diseases, for example, avian pox, are transmitted by "mosquitos" more generally [7]. All related organism information, even associations at varying levels of resolution, need to be clearly described in the ontology.

- *Flag items of biosurveillance relevance to particular sub-fields:* Within biosurveillance, resources focus on paritcular subsets of disease. Some, for example focus on bioterrorism (e.g., BioALIRT [8]), while others focus on reportable diseases (e.g., 122 Cities Mortality Reporting System [9]). In order to maximize utility, we wanted to be able to aggregate diseases that fell under particular categories, as well as diseases that fell within multiple categories. Specifically, we were interested in bioterrorism diseases, diseases of economic importance, US reportable diseases, vaccine-preventable diseases, zoonotic diseases, drug resistant diseases, and emerging or re-emerging diseases.

However, we also wanted to be able to broaden the scope in the future if needed.

- *Specify disease information in varying levels of detail:* Much of biosurveillance takes place under an umbrella of syndromic surveillance [9]. Such systems look for particular clinical symptoms, or syndromes, rather than for confirmed diagnosis of particular diseases. Thus, it was also important that we be able to 1) represent syndromes in the same fashion as diseases and 2) understand the links between syndromes and diseases.

- *Be extensible:* It became clear early on that any ontology for biosurveillance would need to be easily extensible to other data, and potentially to other languages. For example, it is unlikely that our ontology would be comprehensive enough to meet all biosurveillance needs. Thus, the goal was to provide a framework that was simple and useful enough to extend in other directions as it became necessary. We also noted that, while our team works exclusively in English, many in the field of biosurveillance do not. Further, disease names and synonyms change with language. It was important that the ontology we designed be able to extend to other languages for this reason.

- *Be transparent:* Because information about some diseases may be contested (e.g., can Zika be sexually transmitted?), it is imperative that all source documentation be explicit such that users could verify where information came from.

B. *Technical Ontology Requirements*

- *Be easily applied within current biosurveillance tools in different formats (e.g. JSON, OWL):* While OWL is commonly used among ontologists, biosurveillance tools are not typically designed to accept such formats. We wanted to design our ontology in a fashion that would allow export to multiple formats, such that the ontology could be easily applied in different scenarios.

- *Searchable via API:* Related to the above, it was also necessary that internet applications, like those that are common in biosurveillance, have an easy mechanism for use of the ontology. One such mechanism would be an Application Program Interface (API). API's allow other programs to retrieve database results in a simple, computer-readable format. This allows for easy interactions between databases, or between databases and online tools.

C. *Survey of Current Disease Ontologies*

Prior to building our own ontology we surveyed existing ontologies and found two related to disease:

- *Human Disease Ontology/ Disease Ontology:* This ontology was developed as part of the NuGene project and approaches disease ontology from a human, diagnostic perspective [11-12]. Fields focus on anatomy and tissue level processes/ effects of the disease.

- *Infectious Disease Ontology:* This ontology describes infectious diseases but places emphasis on clinical/ laboratory components [13-14]. This emphasis conflicts with our need for both high resolution and generic syndromic categories. This ontology incorporates data from numerous other ontologies including the Human Disease Ontology [14].

In order to understand if these ontologies fit our needs we compared them to the ontology requirements described in IIA. Results of this analysis are given in Table 1.

| Requirements | Disease Ontology | |
|---|---|---|
| | Human Disease Ontology [11-12] | Infectious Disease Ontology [13-14] |
| Synonyms | Yes | No |
| Agents, vectors, & populations associated with diseases described | Only human diseases are described. | Yes |
| Disease properties flagged | Some properties are flagged, but none are biosurveillance related | Some properties are flagged. A few (e.g. drug resistance) overlap with our categories of interest. |
| Disease hierarchy & Syndromic disease categories | Syndromes don't correspond to syndromic biosurveillance. | A hierarchy exists, but there are no syndromic categories |
| Extensible formats | Only OWL or OBO | Only OWL or OBO |

Table 1: The Human Disease Ontology and the Infectious Disease Ontology are compared with respect to the specific needs of a biosurveillance ontology. Requirements on the right are abbreviations of the bullet points within IIA. While both ontologies meet some requires, neither meets all.

The two existing ontologies departed substantially from the comparatively simple description of diseases and relations required for biosurveillance. Neither provided needed properties, syndromes, or a disease hierarchy relevant to biosurveillance. Further, neither was available in a format that was easily "plugged in" to our existing biosurveillance tools. We therefore began development of a biosurveillance relevant ontology.

III. ONTOLOGY STRUCTURE

A. *Definitions*

The following definitions were used to describe ontology fields:

- *Agent:* This is the causative agent or organism of the disease. For example, *Plasmodium vivax* is a causative agent of malaria.

- *Population:* This is the population the disease affects. For example, malaria affects humans.

- *Disease synonym:* These are names referring to the same disease. For example, malaria is sometimes referred to as "Malignant tertian fever".

- *Vector:* This is an organism that helps transmit the disease. It is only present in vector-borne diseases. In the case of malaria, the vector is the Anopheles mosquito.

- *Property:* These are flags of biosurveillance relevance. Malaria is flagged as drug resistant, emerging or re-emerging and US notifiable disease.
- *Transmission:* This is the mechanism for transmission of the disease from one population member to another. Options are binned into air-borne, casual contact, fomite, ingestion, in-utero, sexual transmission, vector-borne and water-borne.
- *Disease parent:* The hierarchical disease parent of a child. Malaria, for example, is a child of the syndrome 'respiratory diseases'.

B. *Schema*

The current schema of our ontology is in Figure 1.

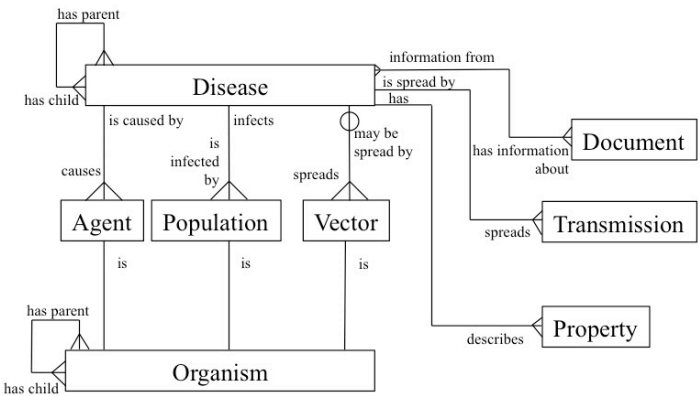

Fig. 1. *Entity relationship diagram for Anthology of Biosurveillance Diseases* Disease has 6 main descriptors: agent, population, vector, property, tranmsission and document. Organisms (agents, populations and vectors) are described by common and scientific names and include a heirarchical component. Transmission and property are catgorical lists with relevant terms and associated descriptions. Document describes source information. Diseases are described by their 6 components as well as through their disease heirarchy.

We used the following methods to address each requirement established above:

- *Correctly identify diseases from synonyms:* Synonyms were curated manually starting with a the base set of synonyms identified by our collaborators at Digital Infuzion, Inc. They started with U.S. notifiable diseases and the human disease subset of Disease Ontology. Synonyms were identified using WordNet [15]. Our team then verified and expanded the disease and synonym lists through literature reviews.

- *Describe disease agents, disease vectors, and populations affected by disease at varying levels of resolution:* All components associated with the organism table have their own fields and heirarchy (see figure 2). Within the organism table we specify parent organisms (parents are heirarchical relationship allowing us to specify that "Anopheles" is a child of "Mosquito"), common name, synonyms, and scientific name. This enables the user to find all diseases that meet criteria at varying levels of resolution. For example, in the user interface, a user can search for diseases that are spread by "insects", "mosquitos", or "Anopheles" specifically. This capability is currently being added to the API. Wherever possible we include the highest resolution reported in literature. In many cases this is species, strain, or serotype information.

- *Flag items of biosurveillance relevance (e.g. bioterrorism diseases):* We selected flags based on categories experts in the field were interested in noting. They include: select agents and toxins, diseases of economic importance, US reportable diseases, vaccine-preventable diseases, zoonotic diseases, drug resistant diseases, and emerging or re-emerging diseases.

- *Specify disease information in varying levels of detail:* Just as organisms have self referencing ties allowing a heirarchy, diseases also have parents. Our disease heirarchy has two components. The first is that some clinical diseases are parents of other diseases. For example, influenza is a parent of avian influenza. The second is that diseases also fall into syndromic categories that are treated like diseases, but are flagged as syndromes. Influenza, in this case, is also a child of "respiratory diseases". The parent to child relationship is a many-to-many one, meaning that diseases can be the children of multiple parents, and visa versa. We believe that this is a very important component of our ontology, because it allows for broader specification of disease than a heirarchy without this component.

We found that, while there are a variety of schemas for describing syndromic disease groups at a high level, there is substantial overlap between them. For the purposes of this ontology we used a modification of the CDC's Essence II categories [16]. Specifically, we use: respiratory, gastrointestinal, febrile, hemorrhagic, dermatologic, and nervous system.

- *Be extensible:* The ontology was originally designed as part of a SQL database. This allows for exceptionally easy addition of new information. For example, we are currently planning to extend the ontology to include relevant ICD-9 and ICD-10 codes. Because of the back end structure, this is a relatively simple process whereby we design a table, connect that table to the disease ontology and add the relevant information. Once this is done, the resulting information is added to the API and can be exported to other formats (e.g., RDF/XML).

- *Be easily applied within current biosurveillance tools in different formats (e.g. JSON, OWL):* The ontology is a SQL database with a Django application overlay. We implemented a REST API using Django's REST API framework that allows users to query the ontology database and export to JSON and XML [17]. Further, we designed an export of the database to RDF/XML compatible with OWL, the format currently utilized by ontologists. This allows computationally experienced users to export in JSON or XML. Our own biosurveillance tools take advantage of the database and the API, while other users can take advantage of other formats (e.g., RDF/XML), as needed.

- *Be transparent:* All references to literature are documented in the database through links to the documents table (see figure 1). Of note, references are *not* currently included in exports, or as part of the API. This feature will be added.

An example of the ontology applied to malaria is given in figure 2. Relationships between organism, agent, population, vector and associations to disease are described, as well as relationships between disease and disease syndrome, and disease and properties/ transmission.

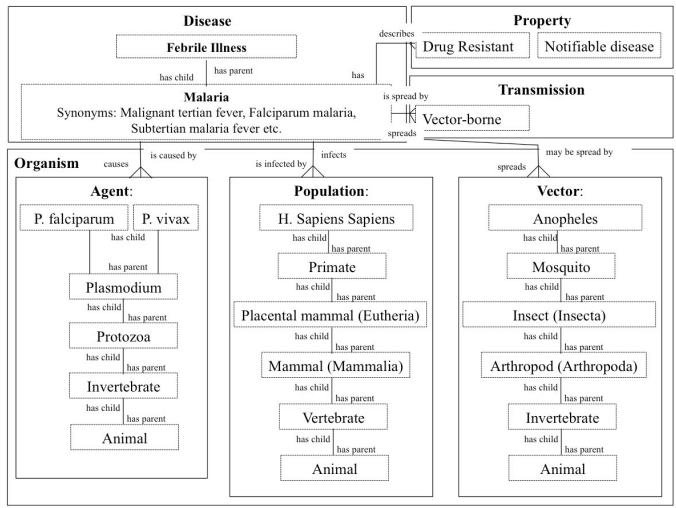

Fig. 2. *Ontology malaria description* Documents have been omitted and some organism associations were truncated for brevity. Both organisms and diseases have heirarchy elements, allowing for optimal searching and more complete disease descriptions. Diseases are described by associated synonyms, properties and transmission.

## IV. FUTURE DIRECTIONS

### A. Applying the ontology to biosurveillance

Using this ontology we have associated specific diseases, or types of diseases, with relevant biosurveillance resources and disease models in the Biosurveillance Resource Directory [18-19]. Additional work of this nature will allow us to facilitate more precise communication between biosurveillance tools currently in development, as well as find and fix flaws in our ontology.

### B. Ontology availability

There is a user interface for the ontology available at: http://brd.bsvgateway.org/disease/
The API is available at: http://brd.bsvgateway.org/api/
The RDF/XML specific export is available at:
http://brd.bsvgateway.org/disease.owlrdf.xml
These URLs may change slightly as our websites grow.

### C. People power vs. computer power

The current process for developing this ontology relies substantially on manual curation by a team of biologists and public health experts. That has allowed us to put a level of detail into the ontology that we believe is beneficial. However, we also recognize the substantial number of hours required to maintain the ontology. We are interested in insights into ways some of this process can be automated.

### D. Next steps

There are some additional features noted above that we hope to add soon: addition of documentation tables to the API, and improvement of API query capabilities. Additional next steps include the set up of a repository for version tracking and to allow outside contributors to make suggestions for content. We believe a community effort for the maintenance of this tool will improve the content and breadth overall.

ACKNOWLEDGMENT

Digital Infuzion, Inc. provided a beginning list of synonyms for some diseases.

We would also like to thank the Defense Threat Reduction Agency (DTRA) for funding this work.

Data Streams Useful for Biosurveillance." *PLOS ONE*, vol 9(1), pp. e83730, 2014.

[19] Available at: brd.bsvgateway.org